\documentclass[aps,floatfix,twocolumn,showpacs]{revtex4-1}
\usepackage{amsmath}
\usepackage{amsfonts}
\usepackage{amssymb}
\usepackage{graphics,epsfig}
\usepackage{graphicx}

\begin{document}

\title{Central density cusps in the Lema\^{i}tre-Tolman solutions}
\author{Kayll Lake}
\email{lake@astro.queensu.ca}
\affiliation{Department of Physics, Queen's University, Kingston,
Ontario, Canada, K7L 3N6 }
\date{\today}

\begin{abstract}
The character of the central density profile in the
Lema\^{i}tre-Tolman (LT) solutions plays a fundamental role in
their application as cosmological models. This same character is
studied here for these solutions used to model complete
gravitational collapse. A necessary condition for the development
of a black hole (not even locally naked singularities) is
developed. This condition allows a finite (invariantly) defined
range in central density cusps. If one demands no density cusps in
the initial conditions, then this work shows that the LT solutions
never produce even locally naked singularities.
\end{abstract}

\pacs{04.20.Cv, 04.20.Dw, 04.20.Jb}

\maketitle

\section{Introduction}
Certainly the most widely used exact solution of the Einstein
equations is that of spherically symmetric inhomogeneous dust,
often referred to as the Lema\^{i}tre-Tolman (LT) (and sometimes
as the Lema\^{i}tre-Tolman-Bondi (LTB)) model. One can find very
detailed discussions of these solutions in some modern texts
\cite{pk}. There is a very extensive application of these models
in cosmology \cite{Bolejko}, and their use in the study of nakedly
singular gravitational collapse goes back at least 35 years
\cite{es}. For general discussions of these models see \cite{pk},
the earlier text \cite{Krasinski}, and \cite{gp}. There are many more recent discussions.
The evolution of radial profiles (which is not of primary concern here) see \cite{sus1},
and for various considerations of gravitational entropy (which are of interest here) see \cite{sus2}.

One of the most interesting applications of the LT models in
cosmology is the reproduction of observables of the $\Lambda$CDM
model without $\Lambda$. The LT models that do this have central
density cusps \cite{ce}. Naturally, such behavior elicits two
points of view: the density profiles are unphysical \cite{vfw},
and the density profiles are just fine \cite{ok}.

The purpose of the present communication is to examine the role
that the central density profiles play in gravitational collapse.
Whereas the usual treatment of the LT models involves coordinates
$(r,\theta,\phi,t)$, where $r$ is some radial coordinate, $\theta$
and $\phi$ are the usual angular coordinates, and $t$ is the
proper time along the geodesic streamlines of the fluid, it is
necessary for the present discussion (as explained below) to
switch to coordinates $(m,\theta,\phi,t)$, where $m$ is the
effective gravitation mass \cite{hm}. For clarity, the solution is
developed from first principles in the next section (see also \cite{pk}).

\section{The LT model}

Starting with Einstein's equations \cite{notation}
\begin{equation}\label{einstein}
    G_{\alpha \beta}=8 \pi T_{\alpha \beta}=8 \pi \rho \; u_{\alpha}u_{\beta},
\end{equation}
where the $u^{\alpha}$ are tangent to the generators of the
geodesic flow, we consider only positive definite energy densities
$\rho>0$, and use comoving synchronous coordinates so that
\begin{equation}\label{metric}
ds^2=e^{\alpha(m,t)}dm^2+R^2(m,t)d\Omega^2-dt^2
\end{equation}
where $d\Omega^2$ is the metric of a unit two-sphere, which we
write in the usual form $d\theta^2+\sin^2(\theta)d\phi^2$, and we
assume the existence of  an origin defined by $R(0,t)=0$ (and all
$t$ derivatives of $R(0,t)=0$). The generators of the flow are
$u^{\alpha}=\delta^{\alpha}_{t}$ and the radial normals are
$n^{\alpha}=\pm e^{-\alpha/2} \delta^{\alpha}_{m}$ so that
$-u^{\alpha}u_{\alpha}=n^{\alpha}n_{\alpha}=1$ and
$u^{\alpha}n_{\alpha}=0$. From
\begin{equation}\label{Gun}
    G_{\alpha \beta}u^{\alpha}n^{\beta}=0
\end{equation}
we find
\begin{equation}\label{first}
    e^{\alpha}=\frac{(R^{'})^2}{1+2E(m)},
\end{equation}
where $E$ is an arbitrary function ($>-1/2$). For convenience, take
\begin{equation}\label{emass}
\mathcal{M} \equiv \frac{R^3}{2}\mathcal{R}_{\theta \phi}^{\;\;\;\;\theta
\phi},
\end{equation}
where $\mathcal{R}$ is the Riemann tensor and so $\mathcal{M}$ is the (invariantly defined)
effective gravitational mass \cite{hm}. We obtain $m=\mathcal{M}$  \cite{gauge}, with $m$ given by
\begin{equation}\label{rdot}
   \dot{R}^2=2(E+\frac{m}{R}).
\end{equation}
To solve Einstein's equations we integrate (\ref{rdot}) (see below).

The LT solutions have two independent invariants derivable from
the Riemann tensor without differentiation. These can be taken to
be
\begin{equation}\label{Ricci}
    \mathcal{R}=8 \pi \rho
\end{equation}
and
\begin{equation}\label{Weyl}
   w=\frac{2^4}{3}\left(4 \pi \rho-\frac{3m}{R^3}\right)^2,
\end{equation}
where $\mathcal{R}$ is the Ricci scalar and $w$ is the first Weyl
invariant ($C_{\alpha \beta \gamma \delta}C^{\alpha \beta \gamma
\delta}$ where $C_{\alpha \beta \gamma \delta}$ is the Weyl
tensor).

From (\ref{rdot}) and (\ref{Ricci}) we arrive at
\begin{equation}\label{rho}
    4 \pi \rho(m,t) = \frac{1}{R^2R^{'}} .
\end{equation}
From (\ref{rho}) we have
\begin{equation}\label{limit}
    \lim_{m \rightarrow 0}\frac{R^3}{m}=\lim_{m \rightarrow 0}3R^2R^{'}=\frac{3}{4 \pi \rho(0,t)}.
\end{equation}
Further, it follows immediately from (\ref{rdot}) (assuming, of
course, some non-vanishing interval in $t$ such that $e^{\alpha}
\neq 0$) that
\begin{equation}\label{ecenter}
    \lim_{m \rightarrow 0}E=0.
\end{equation}
We are interested in the avoidance of naked singularities, and
since these can only arise at $m=0$ \cite{lake}, we take $E=0$, and consider $E(m) \neq 0$ an inessential complication to the considerations presented here \cite{E}.
In the cosmological context, $E(m) \neq 0$ is an essential consideration.
\bigskip

It is clear from (\ref{Ricci}) and (\ref{rho}) that scalar
polynomial singularities occur for
\begin{equation}\label{sing}
    R^2R^{'}=0.
\end{equation}
 A ``bang" (or ``crunch") occurs for $R=0$. Shell crossing
singularities occur for $R^{'}=0$. The conditions for their
avoidance are well known. See \cite{Hellaby1} and \cite{sus1}.

For explicit expressions we now integrate (\ref{rdot}) with $E=0$ to obtain
\begin{equation}\label{re}
   R=(\frac{9m}{2})^{1/3}[t-T(m)]^{2/3},
\end{equation}
and so
\begin{equation}\label{gmm}
    e^{\alpha}=\frac{(t-T-2mT^{'})^2}{[6m^2(t-T)]^{2/3}},
\end{equation}

\begin{equation}\label{rho1}
    2 \pi \rho=\frac{1}{3(t-T)(t-T-2mT^{'})},
\end{equation}
and
\begin{equation}\label{w1}
   w = \frac{2^8}{3^3}\frac{(mT')^2}{(t-T)^4(t-T-2mT')^2}.
\end{equation}
Now $t$ has the freedom of a linear transformation and we restrict
part of that freedom by setting  $T(0)=0$.

\section{Gravitational Collapse}

We have
\begin{equation}\label{rho2}
    2 \pi \rho(0,t)=\frac{1}{3t^2}.
\end{equation}
We take $t$ increasing to the future. The model is non-singular for $t<0$. The
singularity ($s$) starts at $m=t=0$ and propagates out to larger $m$
according to
\begin{equation}\label{s}
    t_{s}=T.
\end{equation}
Shell crossing singularities ($sc$) start at $m=t=0$ and propagate out to larger $m$
according to
\begin{equation}\label{sc}
    t_{sc}=T+2mT'.
\end{equation}
To ensure that $t_{sc} > t_{s}$ for $m>0$ we take $T'>0$ for $m>0$ and so the streamlines of constant $m$, which cannot be propagated through $t_{s}$, never reach $t_{sc}$ for $m>0$.
To ensure that $\rho'<0$ for $m>0$ we need
\begin{equation}\label{negrho}
    t<T+\frac{mT'^2}{2T'+mT''}
\end{equation}
and so for the models considered here $\rho'<0$ everywhere for $m>0$ as long as
\begin{equation}\label{negrho1}
   T''>0
\end{equation}
and so $T$ must be concave up for $m>0$.
\section{Visibility of the Singularity}
It is well known that both branches of the radial null geodesics converge for $R<2m$ \cite{pk}. The apparent horizon locus ($ah$) is therefore given by
\begin{equation}\label{ah}
    t_{ah}=T-\frac{4m}{3}.
\end{equation}
Since $t_{s}>t_{ah}$ for $m>0$ the singularity at $t_{s}$ for
$m>0$ is not visible \cite{note}. However, for $m=0$,
$t_{s}=t_{ah}$, and so there exists the possibility that radial
null geodesics propagate from the singularity at $m=t=0$ to larger
$m$. Since $t_{s}'(0) \geq 0$, in order to avoid null geodesics
propagating from $m=t=0$ we need $t_{ah}'(0)<0$. We therefore have
a sufficient local condition for the formation of a black hole
\cite{bh}:
\begin{equation}\label{bh}
    t_{s}'(0)<\frac{4}{3}.
\end{equation}
The sufficient global condition for the global visibility of the singularity at $m=t=0$ is given by \cite {jk}
\begin{equation}\label{global}
    t_{s}'(m)>\frac{26+15 \sqrt{3}}{3}.
\end{equation}
\section{Initial conditions}
From (\ref{rho1}) it follows that
\begin{equation}\label{central}
    2 \pi \lim_{m \rightarrow 0} \frac{\partial \rho}{\partial m}\biggr| _{t<0} = \frac{4}{3} \frac{t_{s}'(0)}{t^3}.
\end{equation}
From (\ref{bh}) and (\ref{central}) then the sufficient condition for the formation of a black hole can be stated as
\begin{equation}\label{sbh}
    2 \pi  \lim_{m \rightarrow 0} \frac{\partial \rho}{\partial m}\biggr| _{t<0} < \left( \frac{4}{3} \right)^2 \frac{1}{t^3}
\end{equation}
and from (\ref{global}) and (\ref{central})
\begin{equation}\label{sns}
    2 \pi  \lim_{m \rightarrow 0} \frac{\partial \rho}{\partial m}\biggr| _{t<0} > \left( \frac{2}{3} \right)^2 (26+15 \sqrt{3}) \frac{1}{t^3}
\end{equation}
is a sufficient condition for the global visibility of the
singularity. Note that because of the freedom that remains in $t$,
(\ref{central})-(\ref{sns}) are indeterminate up to a
multiplication factor $c$, where $c$ is a constant $>0$. This is of
no consequence here as $c$ can be set by explicit choice of $\rho$
in (\ref{rho2}).

Let us now compare the points of view given in \cite{vfw} and in
\cite{ok}. (In \cite{vfw} an extra derivative was taken in order
to obtain the invariant $\square \; \mathcal{R}$, upon which the
arguments are based \cite{arguments}, as an undefined radial
coordinate $r$ (undefined in the sense of a gauge transformation)
was used. This extra derivative is unnecessary here as
$\frac{\partial \rho}{\partial m}$ is already invariantly
defined.) Whereas the physical context here is different, the
basic physical model is the same (by time inversion) and the basic
physical arguments should apply. According to \cite{vfw}, $\lim_{m
\rightarrow 0} \frac{\partial \rho}{\partial m}| _{t<0}$ should be
$0$. This automatically wipes out the entire subject matter of
shell focusing singularities as follows immediately from
(\ref{bh}). The point of view of \cite{ok} would allow the
development of shell focusing singularities, in principle. It is
worth mentioning that in the cosmological context, whereas the LT
model can be used to interpret current observations, there is no
suggestion that the model should be used at early times. In
contrast, in the collapsing counterpart, it would seem
unreasonable to push the model all the way to the singularity,
where all the interest lies, due to the equation of state. Since
we are interested in matters of principle here, this line of
argument will not be pursued.

\section{Use of a coordinate $r$}
The usual starting point for considerations like those given here
is
\begin{equation}\label{metricr}
ds^2=e^{\alpha(r,t)}dr^2+R^2(r,t)d\Omega^2-dt^2.
\end{equation}
At first sight, it would appear that the development given here
(in terms of $m$) is unnecessary. One need only introduce a
suitably smooth transformation $m=m(r)$, which is one way to set
the gauge freedom in (\ref{metricr}). However, the arguments given
here involve two distinct types of relations: relations like
(\ref{central}) which involve derivatives on both sides of the
equation, and relations like (\ref{bh}) which do not. The first
type allow a smooth transformation from $m$ to $r$ as the
independent variable. The latter do not. Let us write $*$ as any
of $s, sc$ or $ah$. Then since
\begin{equation}\label{rder}
    \frac{dt_{*}}{dr}=\frac{dt_{*}}{dm}\frac{dm}{dr},
\end{equation}
any information contained in $dt_{*}/dm$ is lost. For example,
from (\ref{bh}) we have
\begin{equation}\label{bhr}
    \lim_{r \rightarrow 0} \frac{dt_{s}}{dr}=0
\end{equation}
as the sufficient local condition for the formation of a black
hole. Relying on (\ref{bhr}), we might draw the erroneous (and
entropically unfavorable) global conclusion that black holes in
the LT model must have a constant bang time \cite{jm}. In Figure 1,
I construct a simple counterexample to any such claim by
considering $t_{s}=m^2$, a case both \cite{vfw} and \cite{ok}
would accept.\begin{figure}[ht]
\epsfig{file=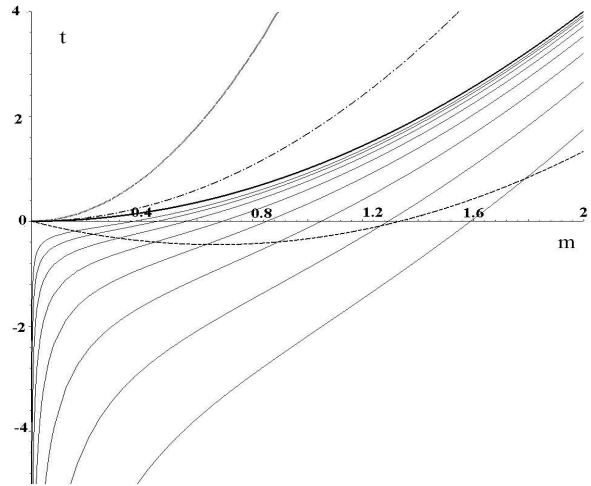,height=3in,width=3.5in,angle=0}
\caption{\label{fig}Complete gravitational collapse to a black
hole for the case $t_{s}=m^2$. The thick solid curve is $t_{s}$,
$t_{sc}=5m^2$ and is shown dotted. $t_{ah}=m^2-4m/3$ and is shown
dashed. $\rho'<0$ for $t<5m^2/3$ shown in the dash dot curve. The
other curves are curves of constant $R$. This shows complete
gravitational collapse to a black hole for a case that has a
variable bang time. Junction can (but need not) be made onto the
Schwarzschild vacuum at any $m>0$. }
\end{figure}
\section{Conclusion}
By using the effective gravitational mass as a coordinate in the
LT solutions, a local sufficient condition for the development of
a black hole has been given. This condition allows and invariantly
defined non-vanishing range in central density cusps, the central
feature of the LT solutions when used to match cosmological
observations without invoking the cosmological constant. If one
demands no density cusps in the initial conditions, then we have
shown that the LT models never produce even locally naked
singularities.
\begin{acknowledgments}
This work was supported in part by a grant from the Natural Sciences and Engineering Research Council of Canada. Portions of this work were
made possible by use of \textit{GRTensorII} \cite{grt}.
\end{acknowledgments}


\begin{thebibliography}{}
\bibitem{notation}We use geometrical units, a signature of $+2$ and designate functional dependence usually only on the first appearance of a function.
Throughout, $^{'}=\partial / \partial m$ and $^{.}=\partial/
\partial t$. Note that no series approximations are made here.
\bibitem{gauge}Because of our choice of gauge, $m$ must increase monotonically away from the origin and so our coordinates do not allow vacuum as a subcase nor can we cover regular maxima as discussed \cite{pk}. Neither limitation is of any concern here.
\bibitem{E} Details associated with the cases $E(m) \neq 0$ will be presented elsewhere.
\bibitem{note}Note that $dt/dm =T^{'}-4/3$ along the $a h$ whereas along any outgoing radial null geodesic $dt/dm=T^{'}+2/3$ when evaluated at the $ah$. The latter is coincident with the tangent to the locus $R=c$ for any constant $c>0$ at the $ah$.
\bibitem{bh}By the term ``black hole" I mean that the singularity at $m=t=0$ is not even locally naked.
\bibitem{arguments} The arguments in \cite{vfw} are based on the claim that $|\square \; \mathcal{R}| \rightarrow \infty$ if there is any cusp in the central density. I am unaware of any published form of $\square \; \mathcal{R}$. It is a complicated object, and difficult to take limits of using a radial coordinate. However, using the coordinate $m$, I find
    \begin{equation*}
        \lim_{m \rightarrow 0} \square \; \mathcal{R}=-\frac{8}{3}\frac{1}{[t-T(0)]^4}.
    \end{equation*}
 The conclusion is that this invariant in fact tells us nothing about density gradients at the origin. Moreover, with our choice $T(0)=0$, the invariant is regular for $t<0$.
\end{thebibliography}
\end{document}